# SOME ASPECTS OF FOUR DIMENSIONAL BLACK HOLE SOLUTIONS IN GAUSS-BONNET EXTENDED STRING GRAVITY

S.O.Alexeyev†and M.V.Sazhin‡

*Sternberg Astronomical Institute, Moscow State University*
*Universitetskii Prospect, 13, Moscow 119899, RUSSIA*
† electronic address: alexeyev@grg2.phys.msu.su
‡ electronic address: sazhin@sai.msu.su

An internal singularity of a string four-dimensional black hole with second order curvature corrections is investigated. A restriction to a minimal size of a neutral black hole is obtained in the frame of the model considered. Vacuum polarization of the surrounding space-time caused by this minimal-size black hole is also discussed.

## 1 Introduction

At the present time the physics of black holes contains a lot of unsolved (and even non-understanded) problems. One of them is the question on the nature of the black hole inner singularities. Studying them we hope to clarify some important aspects of the Cosmic Censorship applications [1,2,3,4,5,6]. Moreover we can also examine the boundaries of the applicability of the General Relativity. Another interesting and completely unsolved question is: what is the endpoint of the black hole evaporation? [2,7,8]. In order to find the more comprehensive solutions of these problems it would be desirable to use the non-minimal gravity model which is the effective low energy limit of some great unification theory. That is why during last years the four-dimensional dilatonic black holes attractes great attention because this type of black holes represents the solution of the string theory at its low energy limit [9,10,11,12,13,14,15,16,17].

It is important to note that the string theory predicts the Einstein equations to be modified by higher order curvature corrections in the range where the curvature of space-time has the near-Planckian values. At the present time the form of the higher order curvature corrections in the string effective action is not investigated completely [15]. We do not know the general structure of the expansion, and, hence, the direct summing up is impossible. But as we deal with the expansion, the most important correction is the second order curvature one which is the product of the Gauss-Bonnet and dilatonic terms. It increases the order of the differential equations till the second one and the existence of the dilatonic term makes the contribution of the Gauss-Bonnet term to be dynamic. So, the action is (for simplicity, only bosonian part is taken into account)

$$\begin{aligned} S &= \frac{1}{16\pi} \int d^4x \sqrt{-g} \bigg[ m_{Pl}^2 \bigg( -R + 2\partial_\mu \phi \partial^\mu \phi \bigg) \\ &- e^{-2\phi} F_{\mu\nu} F^{\mu\nu} + \lambda \bigg( e^{-2\phi} S_{GB} \bigg) \bigg], \end{aligned} \quad (1)$$

$$S_{GB} = R_{ijkl} R^{ijkl} - 4 R_{ij} R^{ij} + R^2.$$

Here $\phi$ is the dilatonic field, $F_{\mu\nu} = q \sin\theta d\theta \wedge d\varphi$ is the Maxwell term and $\lambda$ is the string coupling constant.

The most careful investigation of the discussed model started only a few years ago [9,10,11,12,13,14,15,16,17] and all its predictions are not investigated completely yet. It predicts a change of the solution behavior near singularities or in the regions where the influence of the higher order curvature corrections becomes strong. As the differential equations have a very complicated form, the solutions were obtained by the perturbative [14,17] or numerical [9,10,11,12,13] methods. For example, using these methods a new solution called "neutral Gauss-Bonnet black hole" was found [9,10,11].

The main purpose of our work is to discuss external and internal black hole solutions with dilatonic hair and their main properties. This means that we are interesting in static, spherically symmetric, asymptotically flat solutions providing a regular horizon. Therefore, the most convenient choice of metric (which is usually called as the "curvature gauge") is

$$ds^2 = \Delta dt^2 - \frac{\sigma^2}{\Delta} dr^2 - r^2(d\theta^2 + \sin^2\theta d\varphi^2), \tag{2}$$

where $\Delta = \Delta(r)$, $\sigma = \sigma(r)$. We use this curvature gauge (and the Einstein frame) for more convenient comparison with the Schwarzschild solution.

## 2 Numerical results

For searching the solution in the maximal widest range of the radial coordinate it was necessary to use the most "strong" method for a numerical integration of the systems of the differential equations with particular points. This problem was solved by a modernization of the methods of integration over the additional parameter [11]. This allowed us to investigate the internal structure and the particular points of the black hole using the analysis of the main determinant zeros of the linear system of the differential equations in the non-evident form. So, the system has the following matrix form

$$a_{i1}\Delta'' + a_{i2}\sigma' + a_{i3}\phi'' = b_i, \tag{3}$$

where $i = 1, 2, 3$, matrices $a_{ij}$ $b_i$ are:

$$\begin{aligned}
a_{11} &= 0, \\
a_{12} &= -m_{Pl}^2 \sigma^2 r + 4e^{-2\phi}\lambda\phi'(\sigma^2 - 3\Delta), \\
a_{13} &= 4e^{-2\phi}\lambda\sigma(\Delta - \sigma^2), \\
a_{21} &= m_{Pl}^2 \sigma^3 r + 4e^{-2\phi}\lambda\phi' 2\Delta\sigma, \\
a_{22} &= -m_{Pl}^2 \sigma^2(\Delta' r + 2\Delta) - 4e^{-2\phi}\lambda\phi' 6\Delta\Delta', \\
a_{23} &= 4e^{-2\phi}\lambda 2\Delta\Delta'\sigma, \\
a_{31} &= 4e^{-2\phi}\lambda\sigma(\Delta - \sigma^2), \\
a_{32} &= 2m_{Pl}^2 \sigma^2 \Delta r^2 \phi' + 4e^{-2\phi}\lambda\Delta'(-3\Delta + \sigma^2), \\
a_{33} &= -2m_{Pl}^2 \sigma^2 \Delta r^2 \sigma,
\end{aligned}$$

$$\begin{aligned}
b_1 &= -m_{Pl}^2 \sigma^3 r^2 (\phi')^2 + 4e^{-2\phi}\lambda\sigma(\Delta - \sigma^2) 2(\phi')^2, \\
b_2 &= -m_{Pl}^2 \sigma^3 (2\Delta' + 2\Delta r(\phi')^2) - \frac{1}{2} e^{-2\phi} q^2 \frac{\sigma}{r^3} \\
&\quad + 4e^{-2\phi}\lambda 4\sigma\Delta\Delta'(\phi')^2 - 4e^{-2\phi}\lambda\phi' 2(\Delta')^2 \sigma,
\end{aligned}$$

$$b_3 = 2m_{Pl}^2\sigma^3 r\phi'(\Delta'r+2\Delta) - 4e^{-2\phi}\lambda(\Delta')^2\sigma - 2e^{-2\phi}q^2\frac{\sigma}{r^2}.$$

Matrix (3) represents the linear system of ordinary differential equations (relatively the oldest derivatives) given in a non-evident form. Hence, according to the existence theorem the system (3) has a single solution only in the case of its main determinant to be not equal to zero. In the case of zero main determinant at some point of the solution trajectory, the uniqueness of the solution (3) will be violated. Different types of zeros of the main system determinant correspond to the different types of the particular points of the solution and only three types of zeros present in the asymptotically flat solution of Eqs. (3). So, as the main determinant has the following structure

$$D_{main} = \Delta\left[A\Delta^2 + B\Delta + C\right], \tag{4}$$

where

$$A = (-32)e^{-4\phi}\sigma^2\lambda^2\left[4\sigma^2\phi'^2 m_{Pl}^2 r^2 - \sigma^2 m_{Pl}^2 + 12e^{-2\phi}\Delta'\phi'\lambda\right],$$

$$B = (-32)e^{-2\phi}\sigma^4\lambda\left[\sigma^2\phi' m_{Pl}^4 r^3 + 2e^{-2\phi}\sigma^2\lambda m_{Pl}^2 - 8e^{-4\phi}\Delta'\phi'\lambda^2\right],$$

$$C = 32e^{-4\phi}\sigma^8\lambda^2 m_{Pl}^2 - 2\sigma^8 m_{Pl}^6 r^4$$
$$+ 64e^{-4\phi}\sigma^6\Delta'\lambda^2 m_{Pl}^2 r + 128e^{-6\phi}\sigma^6\Delta'\phi'\lambda^3,$$

zeros of the main determinant are

$$\begin{aligned}(a) &\quad \Delta = 0, \ C \neq 0 \\ (b) &\quad A\Delta^2 + B\Delta + C = 0, \ \Delta \neq 0, \ C \neq 0 \\ (c) &\quad \Delta = 0, \ C = 0.\end{aligned} \tag{5}$$

The numerical results are shown in the Fig. 1. It presents a dependence of the metric functions $\Delta$ (a), $\sigma$ (b) and the dilaton function $e^{-2\phi}$ (c) against the radial coordinate $r$. The solution exist in the definite range of the magnetic charge $q$: $0 \leq q \leq r_h/\sqrt{2}$, which corresponds to the charged solution in the first order[16]. The behavior of the solution outside the horizon has the usual form. Under the influence of the Gauss-Bonnet term the structure of the solution inside the horizon changes such that a new limiting value named "critical magnetic charge" appears. The structure of the inside solution is defnded by this value. When $q_{cr} > q \geq r_h/\sqrt{2}$ the solution has the waiting behavior which corresponds to the fist order one. When $0 \leq q < q_{cr}$ a new singular point so-called $r_s$ appears (see Fig. 1). The type (b) of zero of the main system determinant is realized in this point. The solution turns to the other branch when reaching this point. Only two branches exist near the position $r_s$. There are no any branches between $r_s$ and the origin. Therefore a new singular surface with the topology $S^2 \times R^1$ appears (it is an infinite "tube" in the $t$ direction). Such a singularity is absent in the first order curvature gravity. This singularity exists in the different kinds of metric choice as we tested.

## 3 The couplings $r_s = r_s(r_h, \lambda)$ and $q_{cr} = q_{cr}(r_h, \lambda)$

One can write an asymptotic expansions near the position $r_s$. They are

$$\Delta = d_s + d_1(\sqrt{r-r_s})^2 + d_2(\sqrt{r-r_s})^3 + \ldots,$$

$$\sigma = \sigma_s + \sigma_1\sqrt{r-r_s} + \sigma_2(\sqrt{r-r_s})^2 + \ldots,$$
$$\phi = \phi_s + \phi_1(\sqrt{r-r_s})^2 + \phi_2(\sqrt{r-r_s})^3 + \ldots,$$

where $r - r_s \ll 1$. Using these expansions and the relations between the expansion coefficients one can find the behavior of the Kretchmann scalar $R_{ijkl}R^{ijkl}$ near $r_s$:

$$\begin{aligned}R_{ijkl}R^{ijkl} &= 4\frac{\Delta^2}{\sigma^4 r^4} + 8\frac{\Delta^2(\sigma')^2}{\sigma^6 r^2} - 8\frac{\Delta}{\sigma^2 r^4} - 8\frac{\Delta\Delta'\sigma'}{\sigma^5 r^2} \\ &+ \frac{(\Delta'')^2}{\sigma^4} + 4\frac{(\Delta')^2}{\sigma^4 r^2} - 2\frac{\Delta''\Delta'\sigma'}{\sigma^5} + \frac{(\Delta')^2(\sigma')^2}{\sigma^6} + \frac{4}{r^4} \\ &= \frac{\text{const}}{(r-r_s)} + O\left(\frac{1}{(r-r_s)}\right) \longrightarrow \infty.\end{aligned}$$

All radial time-like and isotropic geodesics formally can be continued under the singularity $r_s$ because they all have the form $\text{Const}*(r-r_s)$. Moreover, using these expansions one can prove that only two branches can exist near the position $r_s$. It can be proved because after the manipulation with the expansions one can obtain only quadratic equation to the coefficient $d_2$.

It is possible to obtain the approximate relations $r_s = r_s(r_h, \lambda)$ and $q_{cr} = q_{cr}(r_h, \lambda)$. Substituting the Schwarzschild expansions of the metric functions (with the vanishing value of the dilaton charge $D$) into the main system determinant and after some manipulations one obtains ($m_{Pl} = 1$)

$$r_s = \lambda^{1/3}\left(4\sqrt{3}r_h\phi_s\right)^{1/3} \tag{6}$$

$$q_{cr} = \lambda^{1/6}\left(\frac{1}{2}\sqrt{3}r_h^4\right)^{1/6}. \tag{7}$$

Fig. 2 (a) and (b) show the graphs of the values $r_s$ and $q_{cr}$ against the coupling parameter $\lambda$ given by these formulas (6-7) and by the numerical integration. From (6) it is possible to conclude that the pure Schwarzschild solution is the limit case of ours with $r_s = 0$. In the case with rather a small value of $\lambda$ these formulas are in good agreement with the results calculated by the numerical integration. While increasing $\lambda$, the absolute error increases as a consequence of ignoring the non-vanishing values (for example $(1-\sigma)$, $\phi'$). It is necessary to point out that these formulas represent the dependencies $r_s = \text{const}\,\lambda^{1/3}$ and $q_{cr} = \text{const}\,\lambda^{1/6}$ which we suppose to be right because after the appropriate selection this constant by hands the agreement between numerical data and this formula improves. Fig. 2 shows also that when the influence of the GB term (or black hole mass) increases, $r_s$ also increases.

## 4  Minimal black hole

### 4.1  Numerical investigations

Figure 3 shows the graph of the metric function $\Delta$ versus the radial coordinate $r$ at the different values of the event horizon $r_h$ when $q = 0$. The curve (a) represents the case where $r_h$ is rather large and is equal to 30.0 Planck unit values (P.u.v.). The curve (b) shows the changes in the behavior of $\Delta(r)$ when $r_h$ is equal to 7.5

P.u.v. The curve (c) represents the boundary case with $r_h = r_{hmin}$ where all the particular points merge and the internal structure disappears. The curve (d) shows the case where $2M \ll r_{hmin}$ and any horizon is absent. Here it is necessary to note that for minimal and near-minimal values of $r_h$ the metric functions can be approximated by the following formulas

$$\Delta = 1 - \frac{r_h}{r}, \tag{8}$$
$$\sigma = 1 - \frac{s_h}{r^8},$$

where $s_h = s_h(r_h)$.

## 4.2 Analytical investigations

One can also obtain the existence of a minimal dilatonic black hole from the analytical manipulations. This type of a system singular point corresponds to the type (5a) of zeros of the main system determinant. The asymptotic expansions near the position $r_h$ have the form

$$\Delta = d_1(r - r_h) + d_2(r - r_h)^2 + \ldots,$$
$$\sigma = s_0 + s_1(r - r_h) + \ldots, \tag{9}$$
$$\phi = \phi_{00} + \phi_1(r - r_h) + \phi_2(r - r_h)^2 + \ldots,$$

where $(r - r_h) \ll 1$.

Substituting the formulas (9) into the system (3), one obtains the following relations between the expansion coefficients ($s_0$, $\phi_0 = e^{-2\phi_{00}}$ and $r_h$ are free independent parameters,

$$d_1(z_1 d_1^2 + z_2 d_1 + z_3) = 0, \tag{10}$$

where:
$$z_1 = 24\lambda^2 \phi_0^2,$$
$$z_2 = -m_{Pl}^4 r_h^3 s_0^2,$$
$$z_3 = m_{Pl}^4 r_h^2 s_0^4,$$

and the parameter $\phi_1$ for $d_1 \neq 0$ is equal to: $\phi_1 = \left[(m_{Pl}^2)/(4\lambda d_1 \phi_0)\right] * [r_h d_1 - s_0^2]$.

When $d_1 = 0$, the metric function $\Delta$ has the double (or higher order) zero. In such a situation the equation for $d_2$ ($d_3$, $d_4$, ...) is a linear algebraic one and there are no asymptotically flat branches.

When $d_1 \neq 0$ the solution of the black hole type takes place only if the discriminant of the equation (10) is greater or equal to zero and, hence, in this case $r_h^2 \geq 4\lambda \phi_0 \sqrt{6}$ ($q = 0$, here we use the Planckian values where $m_{Pl} = 1$). One or two branches occurs and always one of them is asymptotically flat. With the supposition of $\phi_\infty = 0$ (and, as we tested, in this case $1.0 \leq \phi_0 < 2.0$) the **infinum** value of the event horizon is

$$r_h^{inf} = \sqrt{\lambda} \sqrt{4\sqrt{6}}. \tag{11}$$

The analogous formula but in the other interpretation was studied earlier by Kanti et. al.[9]. In the case $q > 0$ such restriction does not exist and regular horizon can take any meaning in the range $[0 \ldots \infty)$.

Consequently, the point $r_{hmin}$ represents the event horizon and the singularity in the same point. One should remember that $\lambda$ is a combination of the fundamental string constants. That is why formula (11) can be reinterpreted as a restriction to the minimal black hole size (mass) in the given model. This restriction appears in the second order curvature gravity and is absent in the minimal Einstein-Schwarzschild gravity.

*4.3 Vacuum polarization*

One of the main feathers of the black hole strong gravitational field is its influence to the structure of the surrounding space-time [8,18,19,20,21,22,23,24,25]. In the semi-classical level this effect can be described by the vacuum polarization and stress-energy tensor expectation values. Our purpose is to compare the $<T_{\mu\nu}>$ of the Schwarzschild black hole with the same value of the string minimal and near-minimal black hole. This work is in progress now. Nevertheless, we would like discuss some interesting preliminary results.

We analyzes the contribution of the massive fields to the vacuum polarization of the string minimal black hole. In this case the contribution will be considerable enough because its mass has the order of the Planck mass $m_{Pl} = \sqrt{\hbar c/G}$. Working almost near the boundaries of applicability, we can study the expansions for Hartle-Hawking vacuum average of the stress-energy tensor. Using Frolov-Zel'nikov expansions [8,18]

$$<T^{(s)}_{\mu\nu}> = -\frac{2}{|g|^{1/2}}\frac{\delta W^{(s)}_{ren}}{\delta g^{\mu\nu}}, \tag{12}$$

which in one-loop approximation reads

$$W^{(s)}_{ren} = \frac{1}{(4\pi m)^2}\frac{1}{8}\frac{1}{7!}\int d^4 x |g|^{1/2} L^{(s)} + O(\epsilon^2), \tag{13}$$

$$\begin{aligned}
L^{(s)} &= a_1 R_{\alpha\beta\gamma\sigma;\epsilon}R^{\alpha\beta\gamma\sigma;\epsilon} + a_2 R^\alpha_\beta R_{\alpha\gamma\sigma\epsilon}R^{\beta\gamma\sigma\epsilon} + a_3 R^{\gamma\sigma}_{\alpha\beta}R^{\epsilon s}_{\gamma\sigma}R_{\epsilon s}R_{\alpha\beta} \\
&+ a_4 R_{\alpha\beta\gamma\sigma}R^{\alpha\beta\gamma\sigma} + a_5 R R_{\alpha\beta}R^{\alpha\beta} + a_6 R^3 + a_7 R_{;\epsilon}R^{;\epsilon} \\
&+ a_8 R_{\alpha\beta;\gamma}R^{\alpha\beta;\gamma} + a_9 R^\beta_\alpha R^\gamma_\beta R^\alpha_\gamma + a_{10} R_{\alpha\beta}R_{\gamma\sigma}R^{\alpha\beta\gamma\sigma}.
\end{aligned}$$

In our coordinates ($\Delta$, $\sigma$, see expansions (8)) the stress-energy tensor average values (12) have the following form (because of the great size of the formulas we show only the $<T^0_0>$ component in the case when the spin of a particle is equal to 1/2)

$$<T^0_0> = T_1/T_2, \tag{14}$$

$$\begin{aligned}
T_1 &= -8775 r^{80} r_h^2 + 9117 r^{79} r_h^3 + \frac{51940}{31} r^{75} r_h s_h \\
&- \frac{53900}{31} r^{74} r_h^2 s_h - 17962560 r^{74} s_h + 108404640 r^{73} r_h s_h \\
&- 179223493 r^{72} r_h^2 s_h + 89133599 r^{71} r_h^3 s_h - \frac{6468}{31} r^{68} s_h^2 \\
&+ \frac{33026}{31} r^{67} r_h s_h^2 - \frac{19208}{31} r^{66} r_h^2 s_h^2 - 583036720 r^{66} s_h^2 \\
&+ 2813006280 r^{65} r_h s_h^2 - 4012428650 r^{64} r_h^2 s_h^2 + 1783041602 r^{63} r_h^3 s_h^2 \\
&+ \frac{32928}{31} r^{60} s_h^3 - \frac{567518}{31} r^{59} r_h s_h^3 + \frac{528220}{31} r^{58} r_h^2 s_h^3
\end{aligned}$$

$$
\begin{aligned}
&- 2729994080r^{58}s_h^3 + 11437564864r^{57}r_hs_h^3 - 14744667622r^{56}r_h^2s_h^3 \\
&+ 6035301470r^{55}r_h^3s_h^3 - \frac{68257}{31}r^{52}s_h^4 + \frac{900522}{31}r^{51}r_hs_h^4 \\
&- \frac{842800}{31}r^{50}r_h^2s_h^4 - 2818841780r^{50}s_h^4 + 10549836302r^{49}r_hs_h^4 \\
&- 12479930891r^{48}r_h^2s_h^4 + 4749439473r^{47}r_h^3s_h^4 + \frac{74039}{31}r^{44}s_h^5 \\
&- \frac{405230}{31}r^{43}r_hs_h^5 + \frac{359660}{31}r^{42}r_h^2s_h^5 - 540258212r^{42}s_h^5 \\
&+ 1847352418r^{41}r_hs_h^5 - 2027148281r^{40}r_h^2s_h^5 + 720283059r^{39}r_h^3s_h^5 \\
&- \frac{46109}{31}r^{36}s_h^6 - \frac{98490}{31}r^{35}r_hs_h^6 + \frac{115640}{31}r^{34}r_h^2s_h^6 \\
&+ 3278996r^{34}s_h^6 - 4899564r^{33}r_hs_h^6 + 1843020r^{32}r_h^2s_h^6 \\
&+ \frac{18963}{31}r^{28}s_h^7 + \frac{86534}{31}r^{27}r_hs_h^7 - \frac{87612}{31}r^{26}r_h^2s_h^7 \\
&- 1820140r^{26}s_h^7 + 3130052r^{25}r_hs_h^7 - 1456108r^{24}r_h^2s_h^7 \\
&- \frac{7203}{31}r^{20}s_h^8 - \frac{882}{31}r^{19}r_hs_h^8 + 47124r^{18}s_h^8 \\
&+ 24438r^{17}r_hs_h^8 + \frac{2597}{31}r^{12}s_h^9 + \frac{98}{31}r^{11}r_hs_h^9 \\
&- 21148r^{10}s_h^9 - 2790r^9r_hs_h^9 - \frac{539}{31}r^4s_h^{10} \\
&+ 4796r^2s_h^{10} + \frac{49}{31}r^{-4}s_h^{11} - 436r^{-6}s_h^{11},
\end{aligned}
$$

$$
\begin{aligned}
T_2 &= r^{88} - 11r^{80}s_h + 55r^{72}s_h^2 - 165r^{64}s_h^3 \\
&+ 330r^{56}s_h^4 - 462r^{48}s_h^5 + 462r^{40}s_h^6 - 330r^{32}s_h^7 \\
&+ 165r^{24}s_h^8 - 55r^{16}s_h^9 + 11r^8s_h^{10} - s_h^{11}.
\end{aligned}
$$

Here one can see that when $s_h = 0$, then $<T_0^0>$ takes its Schwarzschild value

$$
<T_\nu^\mu>_H^{(s)} = \frac{M^2}{10080\pi^2 m^2 r^8}\left[A_s p_\nu^\mu + 3(2B_s + A_s)q_\nu^\mu\right],
$$

where

$$
\begin{aligned}
p_t^t &= -15 + 16\tfrac{2M}{r} \quad p_r^r = -3 + 4\tfrac{2M}{r} \quad p_\theta^\theta = p_\varphi^\varphi = 9 - 11\tfrac{2M}{r} \\
q_t^t &= -10 + 11\tfrac{2M}{r} \quad q_r^r = 4 - 3\tfrac{2M}{r} \quad q_\theta^\theta = q_\varphi^\varphi = -12 + 14\tfrac{2M}{r}
\end{aligned}
$$

Comparing our stress-energy tensor with the Schwarzschild one, we see that their numerical values coincide asymptotically at the infinity, but strongly differs in the neighborhood of the event horizon (for the Planckian masses the difference is about 15 orders). Therefore taking into account the terms of the string theory shows its influence to be stronger near the event horizon where the gravitational field is not a week one. So, we arrive to a conclusion that the application of the General Relativity gives a good results at the large distances from this minimal black hole, but in the neighborhood of this object one must use quantum gravity models.

## 5  Discussion and conclusions

From all the discussed numerical and analytical arguments it is necessary to conclude that string theory gives the new knowledge on the black hole space-times. It provides new types of the internal singularities which are absent in the same models of the General Relativity.

The most interesting consequence of the second order curvature corrections is the existence of a minimal black hole. It exists only in the neutral (quasi-Schwarzschild) case. Using Green-Witten-Schwartz [26] formulas (Chapter 13), we can calculate the value of the string coupling constant $\lambda$, therefore, we can find the numerical value of this minimal black hole mass. It has the value about 0.4 $m_{Pl}$ (and, in that case, $s_h = 1/40$). Speculating about this phenomenon (if this object is stable) we can say that in General Relativity the most realistic models describing the black holes in our Universe are Schwarzschild one and Kerr one. Sometimes ago [27,28] it was established that spinning black hole looses its angular momentum during the Hawking radiation. Therefore, our minimal (quasi-Schwarzschild) black hole being the endpoint of the evaporation can represent the relic remnant of the black holes formed during the initial stages of our Universe formation. This is the very interesting problem and it requires the additional investigations.

## Acknowledgments

S.A. would like to thank the Organizing Committee for the financial support. This work was also supported by RFBR travel grant No. 97-02-27719. M.V.S. acknowledgments Center for Cosmoparticle Physics "Cosmion" (Moscow, RUSSIA) for financial support.

Figure 1: Dependence of the metric functions $\Delta$ (a), $\sigma$ (b) and the dilaton function $e^{-2\phi}$ (c) against the radial coordinate $r$ when $0 \leq q \leq q_{cr}$

Figure 2: The dependencies of the values $r_s$ and $q_{cr}$ against the coupling parameter $\lambda$ given by the formulas (6-7) (lines) and by the numerical integration (squares)

Figure 3: The dependence of the metric function $\Delta$ versus the radial coordinate $r$ at the different values of the event horizon $r_h$ when $q = 0$. The curve (a) represents the case where $r_h$ is rather large and is equal to 30.0 Planck unit values (P.u.v.). The curve (b) shows the changes in the behavior of $\Delta(r)$ when $r_h$ is equal to 7.5 P.u.v. The curve (c) represents the boundary case with $r_h = r_{h\,min}$ where all the particular points merge and the internal structure disappears. The curve (d) shows the case where $2M \ll r_{h\,min}$ and any horizon is absent

Figure 4: The dependencies of $|<T_0^0>(r)|$ at string case with the second order curvature corrections (a) and at Schwarzschild (b) case

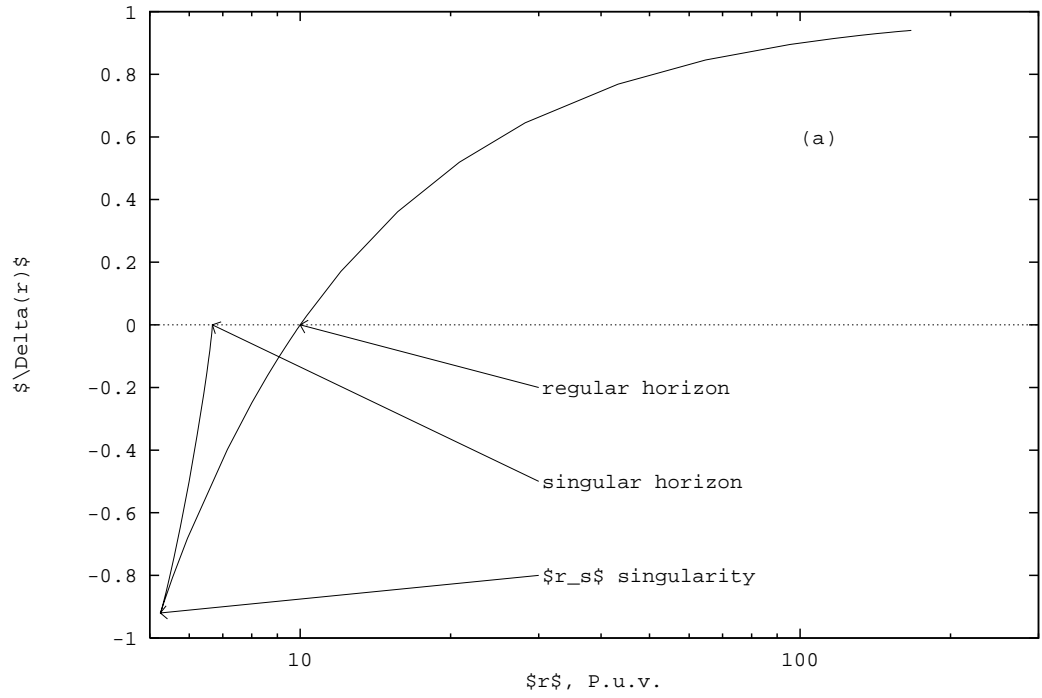

Figure 1: Figure 1a

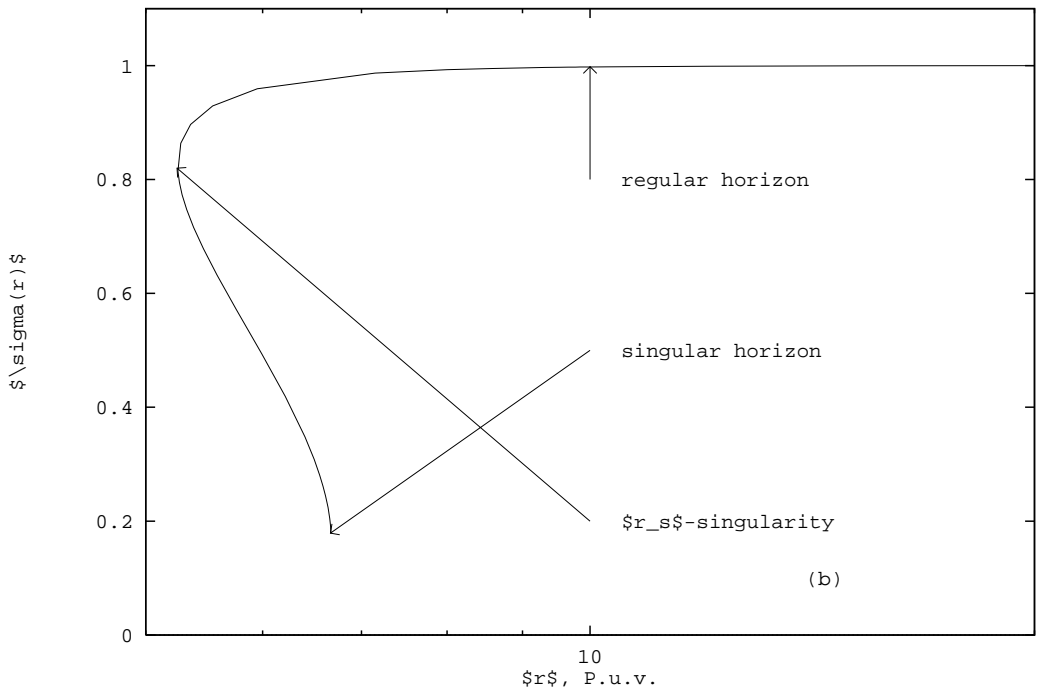

Figure 2: Figure 1b

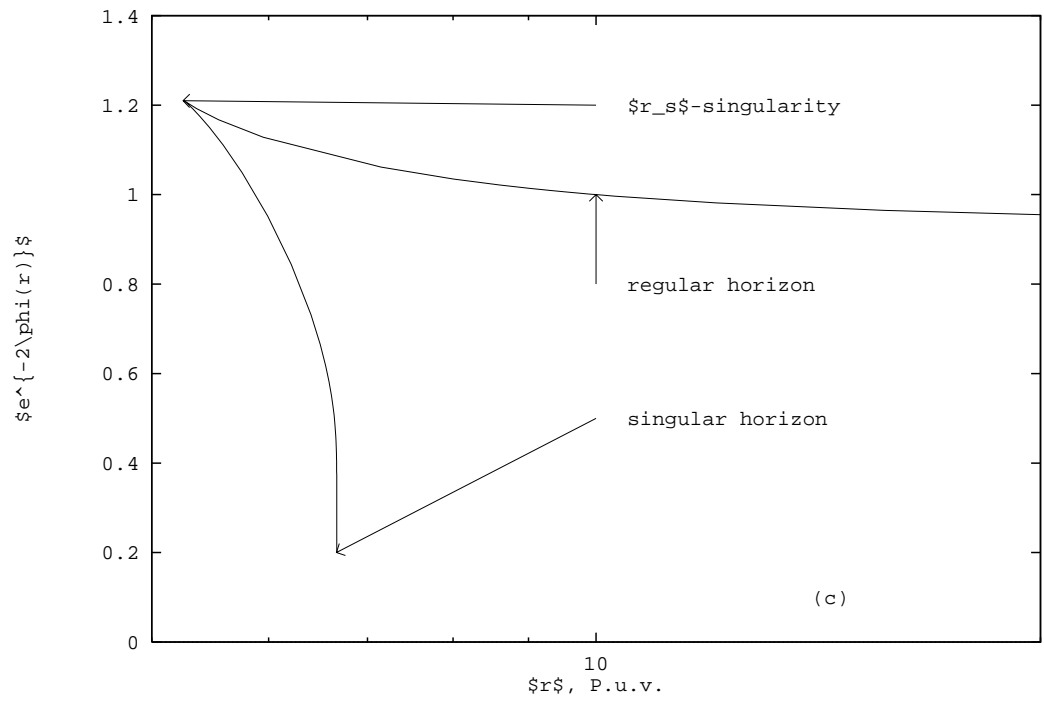

Figure 3: Figure 1c

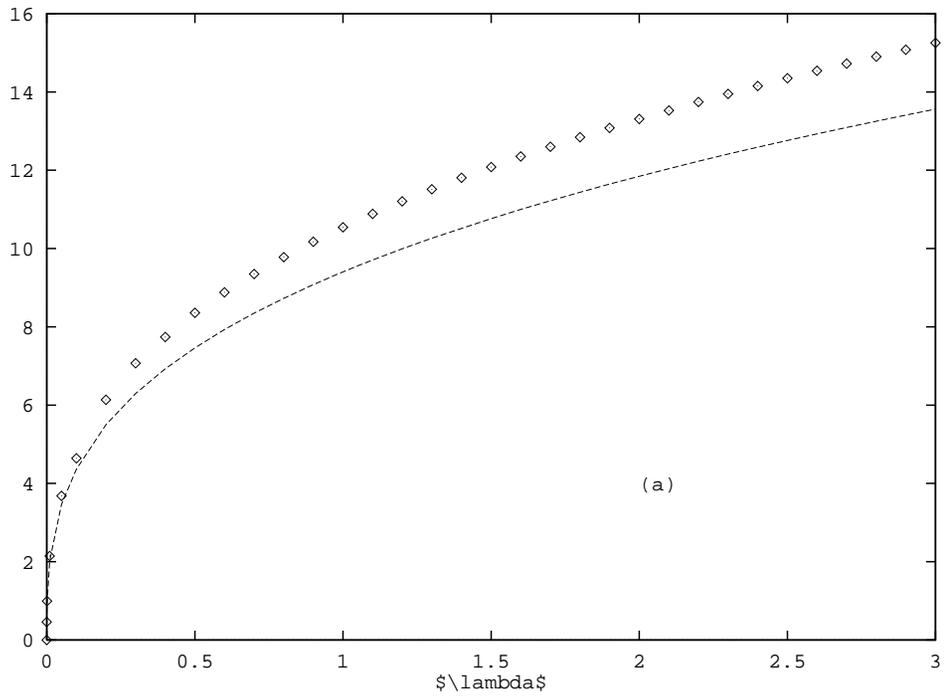

Figure 4: Figure 2a

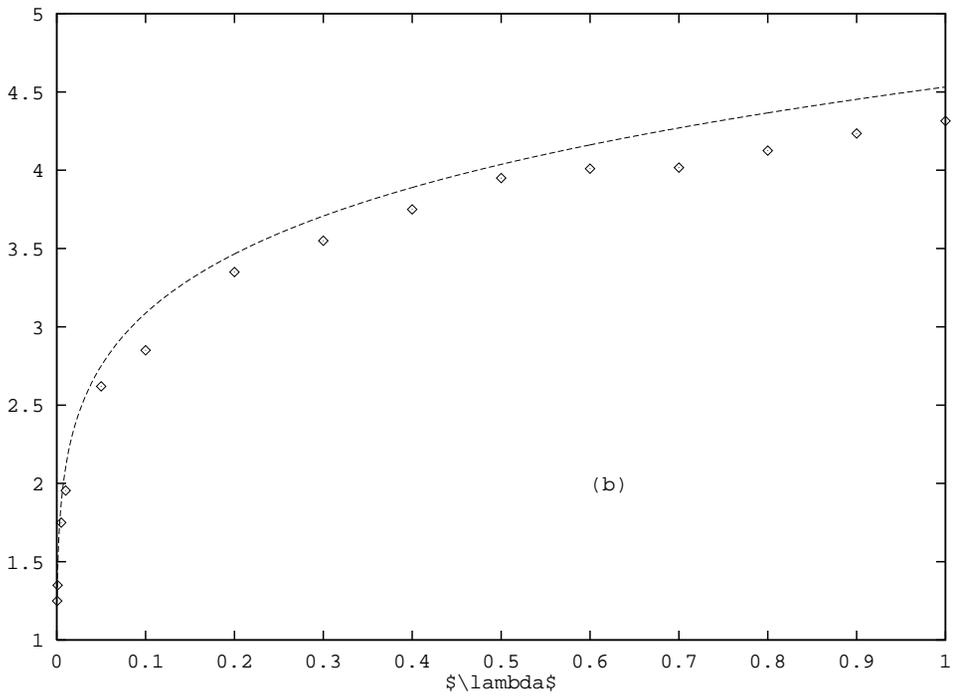

Figure 5: Figure 2b

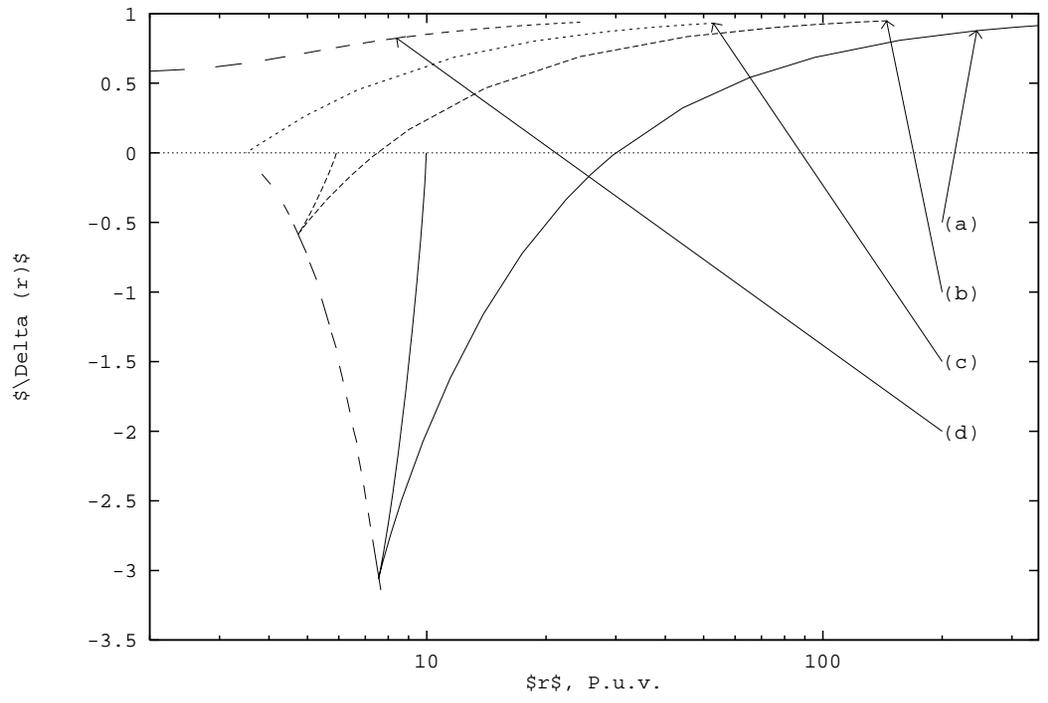

Figure 6: Figure 3

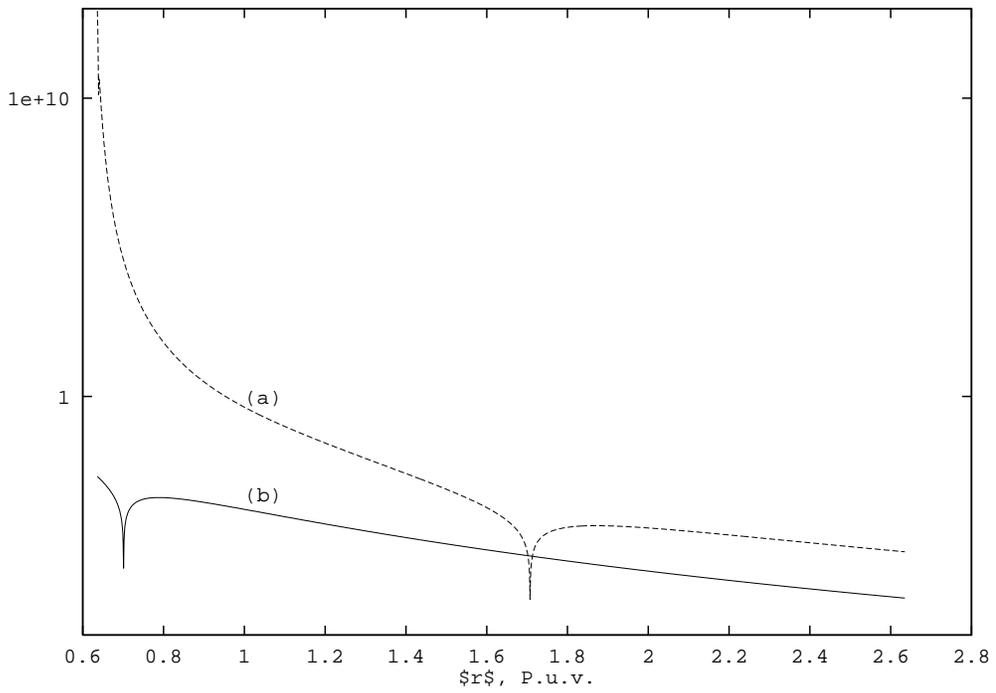

Figure 7: Figure 4